\documentclass[10pt,journal,compsoc]{IEEEtran}
\ifCLASSOPTIONcompsoc
  \usepackage[nocompress]{cite}
\else
  \usepackage{cite}
\fi
\ifCLASSINFOpdf
\else
\fi
\usepackage{amsmath}
\usepackage{algorithmic}
\usepackage{amsmath,amsfonts}
\usepackage{algorithmic}
\usepackage{graphicx}
\usepackage{textcomp}
\usepackage{xcolor}

\usepackage{subfiles} 
\usepackage{multirow}
\usepackage[ruled,linesnumbered]{algorithm2e}
\usepackage{listings}
\usepackage{subcaption}
\usepackage{color,xcolor}
\usepackage{color,xcolor,colortbl}
\usepackage{xspace}
\usepackage{enumitem}
\usepackage{graphicx}
\usepackage{xr}
\usepackage{url}

\usepackage{tcolorbox}

\usepackage{booktabs}
\usepackage{diagbox}
\usepackage{amsmath}
\usepackage{array}

\newtheorem{Def}{Definition}
\newtheorem{Lemma}{Lemma}
\newcommand{\tool}{RECON\xspace}
\newcommand{\eg}{\textit{e}.\textit{g}.\xspace}
\newcommand{\ie}{\textit{i}.\textit{e}.\xspace}

\newcommand{\http}{\url{https://anonymous.4open.science/r/RECON}}

\begin{document}
%


\title{Game Rewards Vulnerabilities: Software Vulnerability Detection with Zero-Sum Game and Prototype Learning}

%
%
%
%

\author{\IEEEauthorblockN{Xin-Cheng Wen\IEEEauthorrefmark{2},
Cuiyun Gao\IEEEauthorrefmark{1}\thanks{* corresponding author.}\IEEEauthorrefmark{2},
Xinchen Wang\IEEEauthorrefmark{2},
Ruiqi Wang\IEEEauthorrefmark{2},
Tao Zhang\IEEEauthorrefmark{4},
and Qing Liao\IEEEauthorrefmark{2}
} \\
\IEEEauthorblockA{\IEEEauthorrefmark{2}}Harbin Institute of Technology, Shenzhen, China\\
\IEEEauthorrefmark{4} Macau University of Science and Technology, Macau, China\\

\IEEEauthorblockA{xiamenwxc@foxmail.com, gaocuiyun@hit.edu.cn, \{200111115, 200111606\}@stu.hit.edu.cn}, tazhang@must.edu.mo, liaoqing@hit.edu.cn}
\IEEEtitleabstractindextext{%
\begin{abstract}

Recent years have witnessed a growing focus on automated software vulnerability detection. Notably, deep learning (DL)-based methods, which employ source code for the implicit acquisition of vulnerability patterns, have demonstrated superior performance compared to other approaches.
However, the DL-based approaches are still hard to capture the vulnerability-related information from the whole code snippet, since the vulnerable parts usually account for only a small proportion. As evidenced by our experiments, the approaches tend to excessively emphasize semantic information, potentially leading to limited vulnerability detection performance in practical scenarios.
First, they cannot well distinguish between the code snippets before (i.e., vulnerable code) and after (i.e., non-vulnerable code) developers' fixes due to the minimal code changes. Besides, substituting user-defined identifiers with placeholders (e.g., "VAR1" and "FUN1") in obvious performance degradation at up to 14.53\% with respect to the F1 score.
To mitigate these issues, we propose to leverage the vulnerable and corresponding fixed code snippets, in which the minimal changes can provide hints about semantic-agnostic features for vulnerability detection.

In this paper, we propose a software vulne\textbf{R}ability d\textbf{E}te\textbf{C}tion framework with zer\textbf{O}-sum game and prototype lear\textbf{N}ing, named \textbf{\tool}. In \tool, we propose a zero-sum game construction module. Distinguishing the vulnerable code from the corresponding fixed code is regarded as one player (\ie Calibrator), while the conventional vulnerability detection is another player (\ie Detector) in the zero-sum game. The goal is to capture the semantic-agnostic features of the first player for enhancing the second player's performance for vulnerability detection. 
To maintain a relative equilibrium between different players while learning the vulnerability patterns, we also propose a class-level prototype learning module for capturing representative vulnerability patterns. The prototype is shared between the detector and calibrator, which captures vulnerability patterns simultaneously.
In addition, to ensure the stability of the training process, we also design a balance gap-based training strategy for the zero-sum game.
Experiments on the public benchmark dataset show that \tool outperforms the state-of-the-art baseline by 6.29\% in F1 score. It can also improve the best baseline by 3.63\% with respect to accuracy in distinguishing between the vulnerable and corresponding fixed code, and 5.75\% in the identifier substitution setting in the F1 score.

\end{abstract}

}

\maketitle

\IEEEdisplaynontitleabstractindextext

%
\IEEEpeerreviewmaketitle

\IEEEraisesectionheading{\section{Introduction}\label{sec:introduction}}
\begin{figure*}[t]
	\centering
	\includegraphics[width=1.0\textwidth]{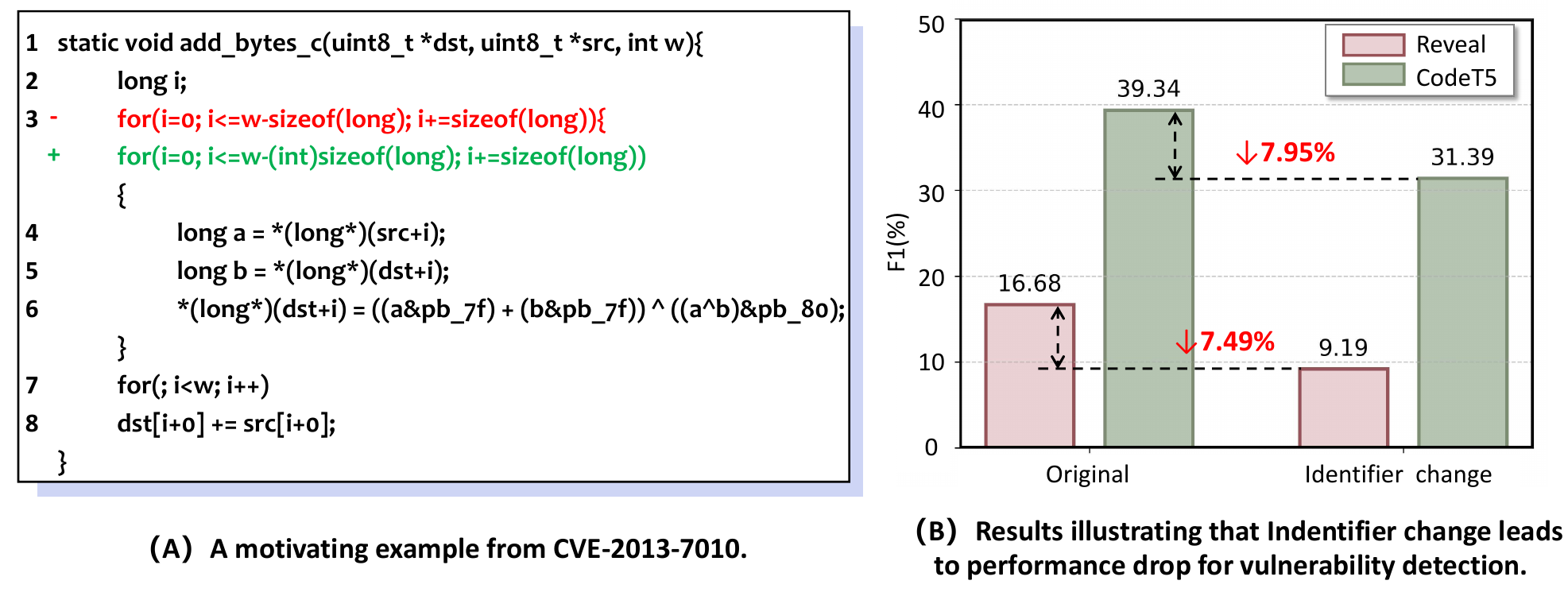}
    \caption{(A) A motivating example from CVE-2013-7010. The red-colored codes are vulnerable code.
    The green-colored codes are the corresponding fixed code.
    (B) The relationship of Reveal~\cite{reveal} and CodeT5~\cite{DBLP:conf/emnlp/0034WJH21/CodeT5} between the original and the identifier change setting. 
    In the dataset with identifiers substituted, we map the identifiers to symbolic names (e.g., VAR1 and FUN1).}
\label{fig:example1}
\end{figure*}
Software vulnerabilities are security issues, which pose substantial security threats and potentially cause significant disruptions within software systems~\cite{DBLP:journals/ieeesp/McGrawP04}. These vulnerabilities represent exploitable weaknesses that, if leveraged by attackers, can lead to breaches or violations of systems'
security protocols, such as cross-site scripting (XSS)~\cite{DBLP:journals/cn/RodriguezTFB20/XSS} or SQL injection~\cite{DBLP:journals/kbs/TangQHLL20/SQL}, is commonly encountered in web applications.
The presence of these vulnerabilities can lead to severe economic consequences. For instance, in 2023, the Clop gang amassed more than \$75 million through MOVEit extortion attacks~\cite{bleepingcomputer}.
Considering the severity
of software vulnerabilities, it is important to develop precise and automated techniques for vulnerability detection.


In recent years, with the increase in the number of software vulnerabilities~\cite{DBLP:journals/pieee/LinWHZX20}, more than 20,000 vulnerabilities have been published each year~\cite{DBLP:conf/uss/HouseholderCNWS20} in the National Vulnerability Database (NVD)~\cite{nvd}. As a result, more and more researchers have focused on automated methods for software vulnerability detection.
We can broadly categorize the existing vulnerability detection approaches into two primary domains: Program Analysis (PA)-based~\cite{DBLP:conf/acsac/ViegaBKM00/2000ITS4/, DBLP:conf/cc/SuiX16/svf, CHECKMARX, DBLP:conf/sp/KimWLO17/VUDDY} and learning-based methodologies~\cite{vuldeepecker, sysevr, DBLP:conf/issta/ChengZ0S22, nguyen2021regvd, cheng2022bug}. PA-based approaches predominantly focus on pre-defined patterns and often rely on human experts to conduct code analysis~\cite{DBLP:journals/virology/LuhMKJS17}. These approaches primarily encompass static analysis~\cite{INFER, DBLP:conf/issta/SuiYX12/saber}, dynamic program analysis~\cite{DBLP:conf/sigsoft/FuRMSYJLS19/fuzz, DBLP:journals/corr/abs-1901-01142/fuzz2}, and symbolic execution~\cite{symbolic, DBLP:conf/IEEEares/LiKBL13/symbolic2}. However, these methods struggle to detect diverse vulnerabilities~\cite{cheng2022bug}.

Conversely, learning-based approaches exhibit greater versatility in detecting various types of vulnerabilities. These approaches typically employ source code or derive program structures from the source code as their input, enabling them to implicitly learn the patterns of vulnerabilities within code snippets. For example, ICVH~\cite{DBLP:conf/ijcnn/NguyenLVMGP21/ICVH} leverages bidirectional Recurrent Neural Networks (RNNs)~\cite{DBLP:journals/tsp/SchusterP97/rnn} with information-theoretic for vulnerability detection. Zhou et al.~\cite{devign} propose Devign, which combines the Abstract Syntax Tree (AST)~\cite{DBLP:conf/icse/ZhangWZ0WL19/ASTNN}, Control Flow Graph (CFG)~\cite{cfg}, Data Flow Graph (DFG)~\cite{Dataflow} and Natural Code Sequence (NCS) into a joint graph and uses the Gated Graph Neural Networks (GGNNs)~\cite{ggnn} to learn the graph representations. CodeBERT~\cite{DBLP:conf/emnlp/FengGTDFGS0LJZ20/codebert} is a Transformer-based pre-trained model and is applied to vulnerability detection via Fine-tuning~\cite{DBLP:journals/tmi/TajbakhshSGHKGL16/finetune}. Although these learning-based methods have made progress in software vulnerability detection, they excessively emphasize semantic information and show limited performance in learning the vulnerability patterns, as evidenced by our experiments.

\textbf{(1) They cannot well distinguish between the code snippets before (i.e., vulnerable code) and after (i.e., non-vulnerable code) developers’ fixes.}
This misclassification occurs because of the minimal code changes between the vulnerable code before and after the fix process.
Fig.~\ref{fig:example1} (A) depicts a motivating example from CVE-2013-7010, which is the vulnerability type of CWE-189 (Numeric Errors)~\cite{CWEID189}. The red-colored statement in line 3 means it will continue iterating as long as ``$i$'' is less than or equal to ``$w - sizeof(long)$''. This may result in out-of-bounds access because it does not consider the case where ``$w$'' is not multiples of ``$sizeof(long)$''. Actually, the green-colored statement (vulnerability-fixed version) only adds one token ``$int$'' to ensure that the loop will not iterate beyond the bounds. This adjustment embodies a subtle yet critical fix, which is hard to be captured by vulnerability detection methods.
For example, CodeBERT~\cite{DBLP:conf/emnlp/FengGTDFGS0LJZ20/codebert} predicts 92.82\% of the code snippets before and after fixes as the same labels, indicating that the method may fail to well learn vulnerability patterns.


\textbf{(2) Substituting user-defined identifiers leads to an obvious performance degradation.}
User-defined identifier names provide rich semantic information of
the source code~\cite{DBLP:journals/infsof/RabinBWYJA21}.
Nonetheless, in most cases, identifier names, such as variable and function names, are irrelevant to the code vulnerability patterns~\cite{DBLP:conf/aaai/ZhangLLMLJ20}. This vulnerability-irrelevant information
can potentially confound vulnerability detection algorithms, leading to erroneous predictions~\cite{DBLP:journals/tosem/ZhangFLMZYSLJ22}.
We also analyze the impact of identifiers on the performance of recent vulnerability detection models such as Reveal~\cite{reveal} and CodeT5~\cite{DBLP:conf/emnlp/0034WJH21/CodeT5}.
Specifically, we create a new dataset with identifiers substituted, in which the identifiers are mapped to symbolic names (e.g., VAR1 and FUN1).
As shown in Fig.~\ref{fig:example1} (B), Reveal and CodeT5 show obvious degradation after the identifier substitution,
dropping 7.49\% and 7.95\%
in terms of the F1 score, respectively.
The results also indicate that the existing methods tend to fail in capturing vulnerability-related information.


To address the above limitations,
we propose a 
software vulne\textbf{R}ability d\textbf{E}te\textbf{C}tion framework with zer\textbf{O}-sum game and prototype lear\textbf{N}ing, named \textbf{\tool}.
 \tool mainly contains two modules:
(1) A zero-sum game construction module, which aims at capturing the semantic-agnostic features for improving the vulnerability detection performance.
Specifically, in the zero-sum game,
one player (\ie Calibrator) is trained to distinguish the vulnerable code from the corresponding fixed code, while another player (\ie Detector) is defined for
the conventional vulnerability detection.
The conventional vulnerability detection means that distinguishes whether the code snippet before the fix contains a vulnerability. 
(2) A class-level prototype learning module for capturing representative vulnerability patterns in each class, which aims to maintain a relative equilibrium in the learning vulnerability patterns of diverse players. The prototype is shared between the detector and calibrator, which captures vulnerability patterns simultaneously.
In addition, to ensure the stability of the training process, we also design a balance gap-based training strategy for the zero-sum game.

We evaluate the effectiveness of \tool{} for software vulnerability detection
in the popular benchmark dataset - Fan et al.~\cite{DBLP:conf/msr/FanL0N20/fan}.
We compare \tool{} with three commonly used GNN-based methods and three pretrained-based methods. The results demonstrate that \tool outperforms all the baseline methods. 
Specifically, \tool{} outperforms 0.94\%, 5.05\%, 7.37\%, and 6.29\% 
in terms of the accuracy, precision, recall and F1 score metrics, respectively. 
It can also improve by 3.63\% with respect to accuracy in distinguishing between the vulnerable and corresponding fixed code, and 5.75\% in the identifier substitution setting in the F1 score.


In summary, the major contributions of this paper are summarized as follows:

\begin{enumerate}

\item We are the first to focus on the zero-sum game to improve software vulnerability detection performance. 

\item We propose \tool, a novel vulnerability detection framework, involves a zero-sum game construction module for capturing the semantic-agnostic features and a class-level prototype learning module for capturing representative vulnerability patterns. We also design a balance gap-based training strategy to ensure the stability of the training process.

\item We perform an evaluation of \tool in the popular benchmark dataset, and the results demonstrate the effectiveness of \tool in software vulnerability detection.
\end{enumerate}

The remaining sections of this paper are organized as follows. 
Section~\ref{sec:background} introduces the background of the zero-sum game and prototype learning.
Section~\ref{sec:architecture} presents the architecture of \tool, which includes two modules: a zero-sum game construction module and a class-level prototype learning module.
Section~\ref{sec:evaluation} describes the experimental setup, including dataset, baselines, and experimental settings. 
Section~\ref{sec:experimental_result} presents the experimental results and analysis.
Section~\ref{sec:discussion} discusses why \tool can effectively detect vulnerability and the threats to validity.  
Section~\ref{sec:conclusion} concludes the paper.

%
%
%
%

\section{Introduction}
\label{sec:introduction}

\section{Background}
\label{sec:background}

\begin{figure*}[t]
	\centering
	\includegraphics[width=0.90\textwidth]{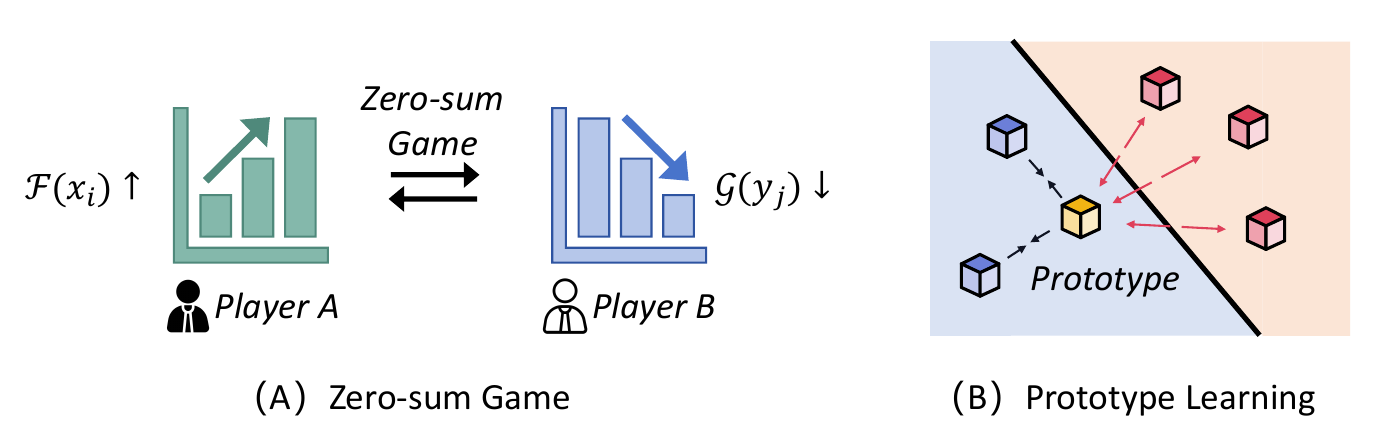}
    \caption{(A) An illustration of the zero-sum game. 
    (B) An example of the prototype learning.}
\label{fig:background_cropped}
\end{figure*}

\subsection{Zero-Sum Game}
\label{sec:backgroundgame}
Game theory~\cite{snidal1985game, chen2006two/game5, DBLP:journals/tac/WuLPLW21/game3, DBLP:journals/tifs/GiboulotPK23/game4, DBLP:conf/aaai/QianWHW23/game2} is the analysis of cooperation and non-cooperation actions within the context of multiple participants, which primarily focuses on the
decision-making process~\cite{kelly2003decision}. In this paper, we restrict our discourse to the simplest case of game theory involving merely two agents, commonly referred to as a \textbf{zero-sum game}~\cite{DBLP:conf/nips/Domingo-EnrichJ20}.
In particular, non-cooperation is more popular due to its more practicality, particularly when considering the presence of competitive dynamics among the participants.
The game between participants can be regarded as a decision-making process, and it can be further regarded as an optimization problem~\cite{DBLP:journals/eswa/XiaoSGL15/optimization}.
Specifically,  
Lemma~\ref{Optimizationproblem} introduces the common 
definition of the zero-sum game:

\begin{Lemma}[Zero-sum game~\cite{shapley1964somegame}]
\label{Optimizationproblem}
Consider that two players are involved in a
zero-sum game. The optimal
solution $x$ of player $A$ and $y$ of player $B$ can be formulated as:

\begin{equation} 
{\mathop{Max}\limits_{x}}\left \lbrace Min\left \lbrace \mathcal{F}(x_{i} \right \rbrace\right \rbrace, \end{equation}

\begin{equation} 
{\mathop{Min}\limits_{y}}\left \lbrace Max\left \lbrace \mathcal{G}(y_{j} \right \rbrace\right \rbrace, \end{equation}

\noindent where
$x = \left[ x_{1}, x_{2},\ldots,x_{n} \right]$ and $y = \left[ y_{1}, y_{2},\ldots,y_{n} \right]$ denote the action sets for the two players, respectively, and each action
is greater than or equal to zero. $\mathcal{F}(x_{i})$ and $\mathcal{G}(y_{j})$ denote the payoff functions of the action $x_i$ and $y_j$, respectively.

\end{Lemma}

Following the definition, we
design the different players and corresponding payoff methods for the vulnerability detection task.
In fact, finding the optimal payoff for both player $A$ and $B$ is a difficult task. As shown in Fig.~\ref{fig:background_cropped} (A), the improvement of one player's payoff function will lead to the corresponding decrease of another player's payoff function. 
It is difficult to maintain a relative equilibrium between different players in the zero-sum game.
In this paper, we construct players A (focusing on before and after the fix) and B (focusing on conventional vulnerability detection) and construct different payoff methods $\mathcal{F}$ and $\mathcal{G}$. Since there is a common goal between the two, i.e., to capture vulnerability-related patterns, we hope to improve vulnerability detection performance through the construction of a zero-sum game. 

\subsection{Prototype Learning}

Prototype learning has gained much attention within pattern recognition in recent years~\cite{DBLP:journals/pieee/Kohonen90/self,DBLP:conf/ijcnn/Kohonen90, DBLP:conf/cvpr/YangZYL18, DBLP:journals/tnn/GevaS91, DBLP:journals/pami/ZhangST23}. One of its most classical and representative methods is the k-nearest-neighbor (KNN) algorithm~\cite{DBLP:journals/ijon/Valero-MasCR16}. It needs to retrieve
the k-nearest neighbors to a sample, effectively forming a local neighborhood. Subsequently, the neighborhood information is leveraged to make a classification decision regarding the sample. 

The key aspect of prototype learning revolves around the strategies for updating the prototypes. 
As shown in Fig.~\ref{fig:background_cropped} (B), 
we have representative prototypes (\ie orange-shaded square) for each class. We need to minimize the distance between samples (\ie blue-shaded squares) and prototypes within the same class while
maximizing the distance between samples (\ie red-shaded squares) and prototypes across different classes simultaneously.
In this paper, we propose a class-level prototype learning module for capturing representative vulnerability patterns in each class.


\section{Proposed Framework}
\label{sec:architecture}
\begin{figure*}[t]
	\centering
	\includegraphics[width=0.99\textwidth]{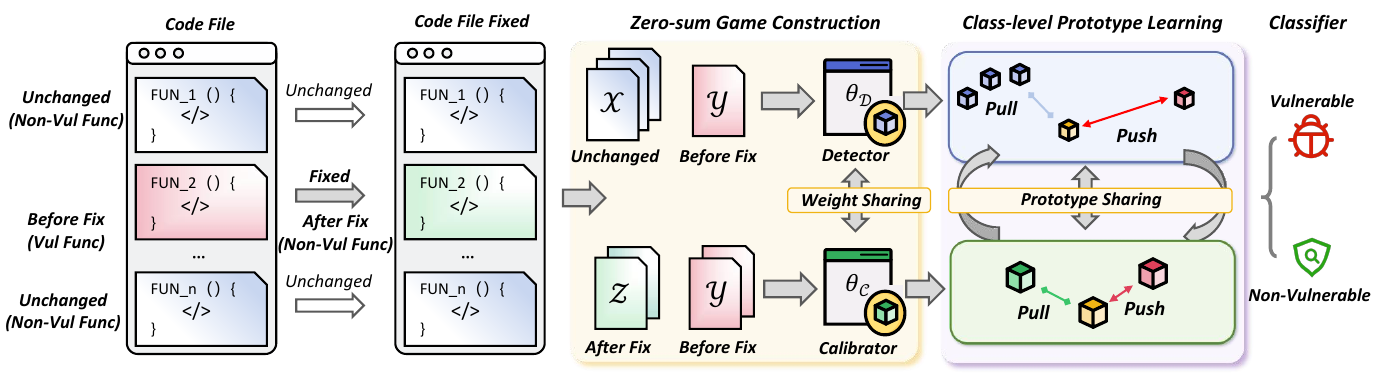}
    \caption{The architecture of \tool. \tool mainly consists of two modules: a zero-sum game construction module and a class-level prototype learning module. 
    The code file and the corresponding fixed code are the input of \tool, including the unchanged code set $\mathcal{X}$, vulnerable code set $\mathcal{Y}$, and vulnerable-fixed code set $\mathcal{Z}$.}
\label{fig:architecture}
\end{figure*}

In this section,
we introduce the overall architecture of \tool. 
As shown in Fig.~\ref{fig:architecture}, \tool mainly consists of two modules: a zero-sum game construction module and a class-level prototype learning module. We also design a balance-gap training strategy for facilitating zero-sum game-based vulnerability detection. We first illustrate our problem formulation and then elaborate on the details of each step in \tool.

\subsection{Problem Formulation}
Similar to the previous studies~\cite{devign, reveal},
the goal of a zero-sum game for software vulnerability detection is to train a binary classifier to determine whether a code snippet contains a vulnerability. 
As shown in Fig.~\ref{fig:architecture},
the sample of data in this paper is formulated as follow,s:
\begin{equation}
(x_{i}, y_{j}, z_{k})|x_{i} \in \mathcal{X}, y_{j} \in \mathcal{Y},  z_{k} \in \mathcal{Z}
\end{equation}
where $\mathcal{X}, \mathcal{Y}, \mathcal{Z}$ denote the unchanged code set, vulnerable code set, and vulnerable-fixed code set, respectively.
$i\in\left\{ {1, 2,..., n_{x}} \right\}, j\in\left\{ {1, 2,..., n_{y}}\right\}, k\in\left\{ {1, 2,..., n_{z}} \right\}$ denote the single sample in different sets, respectively, where $n_{x}$, $n_{y}$, $n_{z}$ denote the numbers of samples in the sets. Each sample $t$ is a pair of source code and the corresponding label (i.e., vulnerable or not), denoted as $<s(t), l(t)>$.
Following the previous methods~\cite{reveal}, we annotate the vulnerable code $y_{j} \in \mathcal{Y}$ as vulnerable and the vulnerable-fixed code $z_{k} \in \mathcal{Z}$ as non-vulnerable. And we annotate all the unchanged code $x_{i} \in \mathcal{X}$ as
non-vulnerable samples.


Based on the input data, \tool aims at learning a mapping $f$ from source code $s(\cdot)$ to its label $l(\cdot)$, i.e., $f: s_{i} \mapsto l_{i}$, to predict whether the given
code snippet is vulnerable or not. The model is trained by a loss function
computed as below:
\begin{equation}
\label{f}
min\sum_{i=1}^{n}\mathcal{L}\left(f\left(s_{i}, {l_{i}}|\left\{s_{i}\right\}\right)\right).
\end{equation}


\subsection{Zero-sum Game Construction}
\label{subsec:zerosum}

\begin{figure*}[t]
	\centering
	\includegraphics[width=0.99\textwidth]{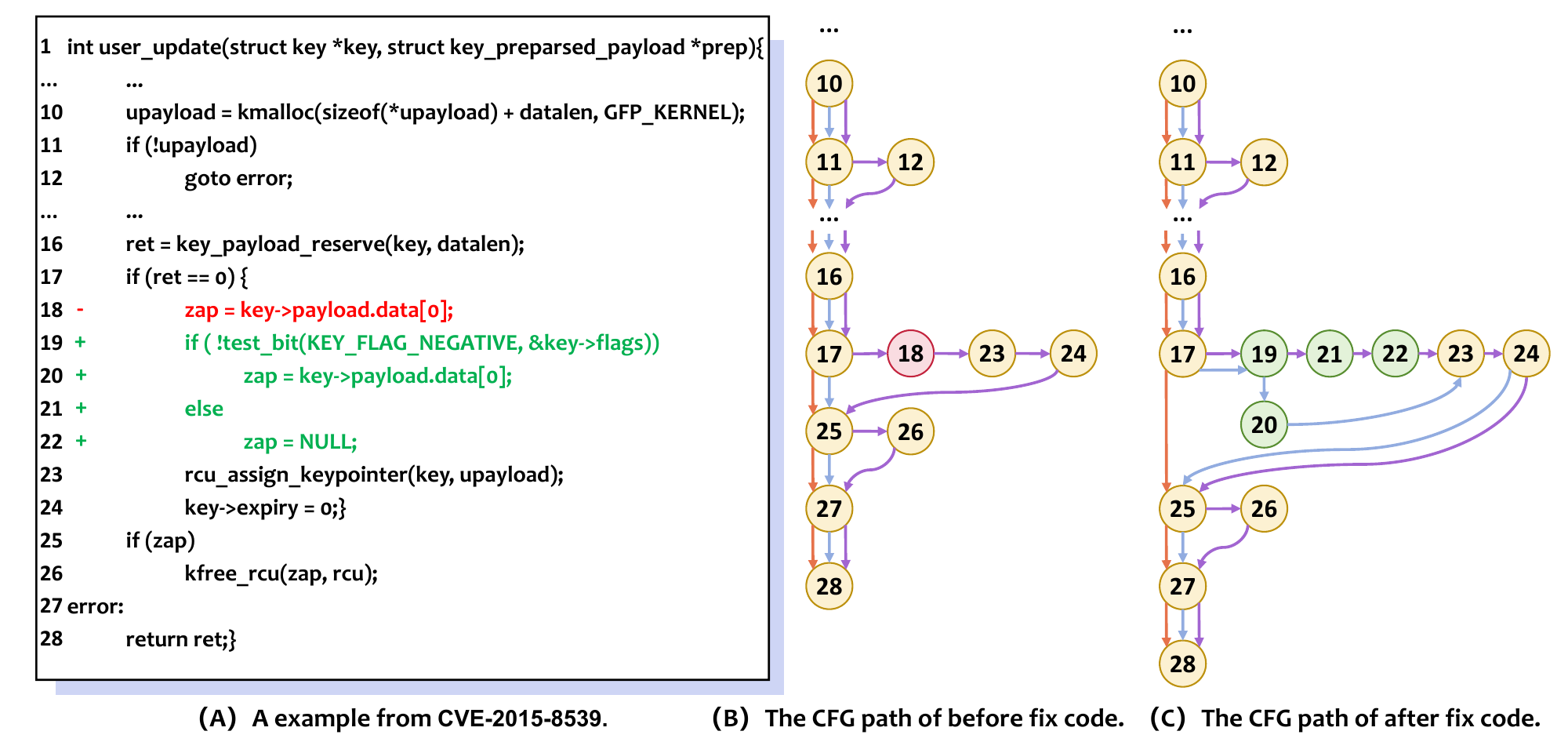}
    \caption{(A) A example from CVE-2015-8539. The red-colored code is vulnerable code. The green-colored code is the corresponding fixed code. (B) The execution path is derived from the Control Flow Graph (CFG) in the vulnerable code. (C) The execution path is derived from the CFG in the corresponding fixed code.}
\label{fig:example2}
\end{figure*}
The zero-sum game construction module aims at leveraging the vulnerable and corresponding fixed code snippets, in which the minimal code changes can provide hints about semantic-agnostic features for vulnerability detection.
As mentioned in Section~\ref{sec:backgroundgame}, we design the two different players: Vulnerability Detector $\mathcal{D}$ and Vulnerability Calibrator $\mathcal{C}$. The module's goal is then transformed to capture
the semantic-agnostic features from
the Calibrator for enhancing the Detector’s performance.

\textbf{(1) Vulnerability Detector $\mathcal{D}$}, which involves the unchanged code set $\mathcal{X}$ and vulnerable code set $\mathcal{Y}$ as the training set. 
Instead of directly using the source code as input, we generate the execution paths derived from the Control Flow Graph (CFG)~\cite{cfg} for each code snippet. 
We use the tree-sitter~\cite{Tree-sitter} to parse the code snippet into CFG.
As shown in Fig.~\ref{fig:example2} (A) and (B), we mark the line number of each statement and encode three distinctive execution paths, which are shaded in orange, blue and purple. Each node represents a line, and the relationship between nodes represents the execution relationship in the source code. 

\textbf{(2) Vulnerability Calibrator $\mathcal{C}$}, which involves vulnerable code set $\mathcal{Y}$ and vulnerable-fixed code set $\mathcal{Z}$ as the training set. 
In contrast to the Detector $\mathcal{D}$, the primary focus of vulnerability calibrator $\mathcal{C}$ lies in discerning alterations 
made to the vulnerable code
$\mathcal{Y}$,
which correspond to the vulnerability fixes.
This strategy is devised to meticulously discern 
the
subtle disparities between code specimens before and after fixes,
specifically in the choice of path differences. As shown in Fig.~\ref{fig:example2} (C), the green circles (including Line 19, Line 21 and Line 22) connected by the purple path correspond to the changes in the fixed code.
We can observe that the differences between execution paths of the vulnerable and fixed code reflect the vulnerability pattern, i.e., 
adding ``$if$'' statements to restrict the addition of keys that already exist but are negatively instantiated in Linux~\cite{CVE-2015}.
By leveraging the execution path, \tool equips the calibrator $\mathcal{C}$ with an effective means of detecting vulnerabilities and distinguishing between the code snippets before and after developers’ fixes.


Then, we use
CodeBERT~\cite{DBLP:conf/emnlp/FengGTDFGS0LJZ20/codebert}
to initialize the representation of each code sample $(I_{i}, l_{i})$.
Specifically, we obtain the CLS token~\cite{DBLP:conf/emnlp/LuHXKMDBLO21/CLS} as the sequence vector $I_{i,j}$ for each execution path in Detector and Calibrator.
We then use TextCNN~\cite{DBLP:journals/tip/HeH0Y16/textcnn} to learn the sample's feature
$s_i$ by involving
the local information:

\begin{equation}
     s_i = MaxPool \left( \left(Concat^{N}_{j=0}(I_{i,j}) \right) \ast W^k \right) || \left(Concat^{N}_{j=0}(I_{i,j}) \right),
\end{equation}
where $Concat(\cdot)$ denotes the combination of vectors of different paths and $\ast$ denotes the convolution operator. $W \in R^{C_{in} \times C_{out} \times k}$ denotes the convolution kernel, where $C_{in}$, $C_{out}$ and $k$ are the input channel, output channel and filter of the convolution, respectively. The symbol $||$ denotes the fusion operator between the convolution results and the sequence vector. $N$ is the number of execution paths extracted from code sample $(I_{i}, l_{i})$.

\subsection{Class-level Prototype Learning Module}

The class-level prototype learning module aims to capture representative vulnerability patterns.
During the model training process, a noteworthy disparity arises in capturing vulnerability features between the Detector $\mathcal{D}$ and Calibrator $\mathcal{C}$, as introduced in Section~\ref{sec:backgroundgame}. Therefore, we need to maintain a relative equilibrium in
learning the vulnerability patterns between Detector $\mathcal{D}$ and Calibrator $\mathcal{C}$, for vulnerability detection.

Specifically, we propose a class-level prototype learning loss, in which
the loss function $\mathcal{L}$ is designed
as below:
\begin{equation}
    \mathcal{L} =  \mathcal{L}_{CE} + \mathcal{L}_{Proto} + \mathcal{L}_{reg},
\end{equation}

\noindent where $\mathcal{L}_{CE}$ is the typical cross-entropy loss for classification, $\mathcal{L}_{Proto}$ is mainly used to mine the relation between the
sample's feature $s_i$ and prototypes, and $\mathcal{L}_{reg}$ is used to improve the generalization performance. 
We leverage the Multi-Layer Perceptron (MLP)~\cite{DBLP:journals/tip/HeH0Y16/mlp} classifier and introduce the traditional $\mathcal{L}_{CE}$:

\begin{equation}
    \mathcal{L}_{CE} = -log(p_{CE}(s_i)),\quad p_{CE}(s_i) = Softmax\left (MLP \left(tanh \left (s_{i} \right) \right) \right).
\end{equation}

The prototypes are
trainable vector representations, denoted as $m_l$ and $l \in C = \{0,1\}$ representing the corresponding label. The prototypes have the similar vector sizes as the
feature $s_i$.
The $\mathcal{L}_{Proto}$ aims at decreasing the distance between the features of all vulnerable samples and prototypes, so that
the prototypes can be regarded as
``vulnerability patterns''.
Specifically, we use the distance $||s_i - m_{l}||^{2}_{2}$ to measure the similarities between sample features and
prototypes.
The prototype loss function $\mathcal{L}_{Proto}$
can be measured as:


\begin{equation}
    \mathcal{L}_{Proto} = -log(p_{Proto}(s_i)),\quad p_{Proto}(s_i) = \frac{e^{-\gamma ||s_i - m_{l}||^{2}_{2}}}{\sum_{j=0}^{C}e^{-\gamma ||s_i - m_{j}||^{2}_{2}}}
\end{equation}

\noindent where $||s_i - m_{l}||^{2}_{2}$ denotes the distance between the feature $s_i$ and prototype $m_l$ in class $l \in C$ and $\gamma$ denotes the hyper-parameter. The $p_{Proto}(s_i)$ indicates the probability of an input feature $s_i$ belonging to the class $l$.
The regularization loss $\mathcal{L}_{reg}$ is defined as below for avoiding over-fitting and improving
the generalization:
\begin{equation}
    \mathcal{L}_{reg} = \lambda||s_i - m_{nearest}||^{2}_{2},
\end{equation}
where $m_{nearest}$ indicates the nearest prototype vector to
the feature $s_i$ in the class $l$, and $\lambda$ is 
a hyper-parameter to control the weight of $\mathcal{L}_{reg}$.

\subsection{Balance Gap-based Training Strategy}
We propose a Balance Gap (BG)-based training strategy to ensure the stability of the training process.
Specifically, \tool can be regarded as a dynamic minimization-vs-minimization game process between Detector $\mathcal{D}$ and Calibrator $\mathcal{C}$ .
This process is formally defined as:
\begin{equation}
\label{minmin}
{\mathop{Min}\limits_{\theta_{\mathcal{D}}}} \left \lbrace {\mathop{Min}\limits_{\theta_{\mathcal{C}}}} \left \lbrace \mathcal{R}(\theta_{\mathcal{D}}, \theta_{\mathcal{C}} \right \rbrace\right \rbrace 
\end{equation}
where Detector $\mathcal{D}$ and Calibrator $\mathcal{C}$ are parametrized by $\theta_{\mathcal{D}}$ and $\theta_{\mathcal{C}}$, respectively. The optimization of the zero-sum game $\mathcal{R}$ is optimized via gradient descent during the training epoch. 
The BG in the zero-sum game is defined as below:
\begin{Def}[Balance Gap]
Considering the training target $\mathcal{R}{(\cdot,\cdot)}$ for the software vulnerability detection,
the previous model parameters $(\theta_{\mathcal{D}},\theta_{\mathcal{C}})$ are updated to  $(\hat{\theta_{\mathcal{D}}},\hat{\theta_{\mathcal{C}}})$ via the gradient descent after each epoch training. Then the balance gap can be formulated as:
\begin{equation}
\label{BG}
BG = \mathcal{L} \left ( \mathcal{R}(\hat{\theta_{\mathcal{D}}},\hat{\theta_{\mathcal{C}}})  \right )- \mathcal{L} \left ( \mathcal{R}(\theta_{\mathcal{D}},\theta_{\mathcal{C}}) 
\right ),
\end{equation}
\end{Def}

\noindent where $\mathcal{L}$ denotes the loss function of \tool. 


The BG fundamentally establishes the relations between the Detector $\mathcal{D}$ and Calibrator $\mathcal{C}$ in the training process, which
consists of two updating steps: 
\textbf{(1) Updating Detector $\mathcal{D}$.} In the first step, we freeze the Calibrator's parameters $\theta_{\mathcal{C}}$ and optimize the Detector $\mathcal{D}$ problem for minimizing the $\mathcal{R}(\theta_{\mathcal{D}}, \theta_{\mathcal{C}})$. 
\textbf{(2) Updating Calibrator $\mathcal{C}$.} Then, in the second step, we freeze the optimal Detector's parameters $\hat{\theta_{\mathcal{D}}}$ and optimize the Calibrator $\mathcal{C}$ problem for minimizing the $\mathcal{R}(\hat{\theta_{\mathcal{D}}}, \theta_{\mathcal{C}})$. 

 

To avoid the large deviation in the training process, we also use the following boundary strategy:
\begin{equation}
\label{boundary}
\Arrowvert BG  \Arrowvert \leq L \Arrowvert \mathcal{L} \left ( \mathcal{R}(\hat{\theta_{\mathcal{D}}},\hat{\theta_{\mathcal{C}}})  \right )- \mathcal{L} \left ( \mathcal{R}(\theta_{\mathcal{D}},\theta_{\mathcal{C}}) 
\right ) \Arrowvert
\end{equation}
where $L$ is the Lipschitz constant~\cite{jones1993lipschitzian}. 
Following the previous proof~\cite{DBLP:conf/aaai/QianWHW23/proof}, 
we can find that the boundary strategy can only be satisfied if $\Arrowvert BG  \Arrowvert = 0$. 
Constrained by the instability of the training process, we propose to train \tool with max-patience~\cite{DBLP:journals/corr/abs-2002-06305/earlystop}.

\section{EXPERIMENTAL Setup}
\label{sec:evaluation}
\subsection{Reasearch Questions}

We evaluate the proposed \tool and aim to answer the following research questions (RQs):

\begin{enumerate}[label=\bfseries RQ\arabic*:,leftmargin=.5in]
    
    \item How does \tool perform compared with the state-of-the-art vulnerability detection approaches?
    \item How effective is \tool in 
    the identifier-substitution setting?
    \item How does \tool perform in distinguishing the vulnerable and the corresponding
    fixed code?
    \item How effective is \tool on 
    the training data
    split by time?
    \item What is the influence of different modules on the detection performance of \tool?

    \item How do the different hyper-parameters impact the performance of \tool?

\end{enumerate}

\subsection{Datasets}
\begin{table*}[t]
\centering
\setlength{\tabcolsep}{5mm}
\renewcommand{\arraystretch}{}

\caption{Statistics of the datasets. The ``Ratio'' indicates the percentage of vulnerable samples relative to non-vulnerable samples.}
\begin{tabular}{c|rrrr}
\toprule

Dataset settings& \multicolumn{1}{c}{\# Vul.} &  \multicolumn{1}{c}{\# Non-Vul.} &\multicolumn{1}{c}{ \# Ratio} & \#Train: \#Valid: \#Test\\

\midrule
Original/Indentifier-substitution/Time-split   &7,238&158074&1: 21.84& 132,229: 16,526 : 16,557   \\
Vulnerability-fix Pair (Pair set)      &  6,998  & 6,998  &1: 1 &11,166: 1,456: 1,374\\
Vulnerability-fix Pair (Combine set)      &7,238&165,072&1: 22.81& 143,395: 17,982 : 17,931   \\
\bottomrule

\end{tabular}
\label{dataset}
\end{table*}

To address the above questions, we use the popular benchmark dataset proposed by Fan et al.~\cite{DBLP:conf/msr/FanL0N20/fan} for software vulnerability detection. This dataset encompasses 91 distinct types of vulnerabilities extracted from 348 open-source GitHub projects. It comprises an extensive repository of approximately 188,000 data samples, including 10,000 vulnerable samples.

We use the Fan et al. dataset because it is the only available vulnerability dataset that provides the corresponding fixed version of the vulnerability code.
In this paper, limited by the text length accepted by CodeBERT~\cite{DBLP:conf/emnlp/FengGTDFGS0LJZ20/codebert}, we removed all samples with lengths greater than 512 tokens to ensure the fairness of the experimental results following the prior study~\cite{DBLP:conf/emnlp/0034WJH21/CodeT5}.


\subsection{Baseline Methods}

In our evaluation, we compare \tool with three 
GNN-based methods and three pre-trained model-based methods.

\begin{enumerate}
\item \textbf{Devign}~\cite{devign}: Devign generates a joint graph by AST, CFG, DFG and Natural Code Sequence (NCS). It then uses a GGNN and a convolution layer for vulnerability detection.

\item \textbf{Reveal}~\cite{reveal}: Reveal constructs the code property graph as input and uses a two-step vulnerability detection framework, including a feature extraction process by GGNN and a representation learning process by the MLP and triplet loss.

\item \textbf{DeepWukong}~\cite{DBLP:journals/tosem/ChengWHXS21/Deepwukong}: DeepWukong uses the DOC2VEC to transform the tokens from source code into node vector representations,
which leverages a graph convolutional network for vulnerability detection.

\item \textbf{CodeBERT}~\cite{DBLP:conf/emnlp/FengGTDFGS0LJZ20/codebert}: CodeBERT is an encoder-only pre-trained model, which is based on Roberta. It is applied to vulnerability detection by fine-tuning.

\item \textbf{CodeT5}~\cite{DBLP:conf/emnlp/0034WJH21/CodeT5}: CodeT5 is a Transformer-based model, which regards the task as the sequence-to-sequence paradigm. It is also applied to vulnerability detection by fine-tuning.


\item \textbf{EPVD}~\cite{DBLP:journals/tse/ZhangLHXL23/EPVD}: EPVD proposes an algorithm for execution path selection.
It adopts a pre-trained model and a convolutional neural network to learn the path representations.

\end{enumerate}

\subsection{Implementation Details}
In all research questions, to ensure the fairness of the experiments, we use the same data splitting for all the approaches.  We try our best to reproduce all baseline methods from publicly available source code,  and use the same hyper-parameters
as their original use.
In this paper, we experiment under
four settings in Fan et al.~\cite{DBLP:conf/msr/FanL0N20/fan} as follows, with
the detailed data statistics
shown in Table~\ref{dataset}: 

\textbf{(1) Original Setting} (in RQ1): Following the previous work~\cite{devign, reveal}, we randomly split the datasets into disjoint training, validation, and test sets in a ratio of 8:1:1. 

\textbf{(2) Identifier-substitution Setting} (in RQ2): In RQ2, we map the user-defined variables and functions to symbolic names in a one-to-one manner (\eg, VAR1 and FUN1) to construct the user-defined identifier-substitution setting. This setting utilizes the same data split as described in RQ1.

\textbf{(3) Vulnerability-fix Pair Setting} (in RQ3): In RQ3, we introduce the pair set comprising pairs of the vulnerable code and
corresponding fixed code (denoted as
``Pair set''). Subsequently, we partition the pair set in an 8:1:1 ratio, with a strict constraint that the vulnerable code and the corresponding fixed code must coexist in the same partition, thus preventing any inadvertent data leakage. Furthermore, we consolidate all code snippets into a combined test set (denoted as
``Combine set''), including samples from both the original and pair set, for comprehensive evaluation purposes.

\textbf{(4) Time-split Setting} (in RQ4): In RQ4, we also utilize a temporal splitting approach for software vulnerability detection. Specifically, we leverage the "update date" as provided in the Fan et al. dataset.  In accordance with this criterion, we divide the data into training, validation, and test sets, maintaining an 8:1:1 ratio while strictly preserving the temporal order of the data points. There is no temporal overlap between these sets, ensuring the integrity of our evaluation process.


We fine-tune the pre-trained model CodeBERT~\cite{DBLP:conf/emnlp/FengGTDFGS0LJZ20/codebert} with a learning rate of $2e-5$. We train our method on a server with 2*NVIDIA A100-SXM4-40GB and the batch size is set to $16$. 
\tool is trained for a maximum of 24 epochs with 5-epoch for max-patience.

\subsection{Performance Metrics}

We employ four widely-used metrics to evaluate the performance of \tool:

\textbf{Precision:} It is defined as $Precision = \frac{TP}{TP + FP}$. Precision quantifies the proportion of true vulnerabilities correctly identified among all the retrieved vulnerabilities. $TP$ represents the number of true positives, while $FP$ represents the number of false positives.

\textbf{Recall}: It is defined as $Recall = \frac{TP}{TP + FN}$. This metric assesses the percentage of actual vulnerable instances that have been successfully identified out of all the vulnerable instances. $TP$ denotes the number of true positives, and $FN$ denotes the number of false negatives.

\textbf{F1 Score}: It is defined as $F1\ score = 2 \times \frac{Precision \times Recall}{Precision + Recall}$. F1 score serves as the harmonic mean of the precision and recall metrics, providing a balanced evaluation of our tool's performance in terms of both precision and recall.

\textbf{Accuracy}: It is defined as $Accuracy = \frac{TP + TN}{TP + TN + FN + FP}$. Accuracy measures the percentage of correctly classified instances out of all instances $TP + TN + FN + FP$.  $TN$ denotes the number of true negatives.



\section{Experimental Results}
\label{sec:experimental_result}
\begin{table*}[htbp]
\centering

\setlength{\tabcolsep}{1.5mm}
\renewcommand{\arraystretch}{1.2}
\caption{
Comparison results between \tool and existing vulnerability detection methods in the original setting (RQ1) and
identifier-substitution setting (RQ2). The best result for each metric is highlighted in bold. 
The cells in grey represent the performance of the top-3 best methods in each metric, with darker colors representing better performance.}
\resizebox{.97\textwidth}{!}{
\begin{tabular}{l|cccc@{\hspace{5mm}}|@{\hspace{5mm}}cccc}
\toprule
\diagbox{Metrics(\%) }{Setting}   & \multicolumn{4}{c@{\hspace{5mm}}|@{\hspace{5mm}}}{RQ1 (Original)}  & \multicolumn{4}{c}{RQ2 (Identifier-substitution)}  \\
\midrule
Method & Accuracy & \multicolumn{1}{l}{Precision} & \multicolumn{1}{l}{Recall} & \multicolumn{1}{l|@{\hspace{5mm}}}{F1 score} & \multicolumn{1}{l}{Accuracy} & \multicolumn{1}{l}{Precision} & \multicolumn{1}{l}{Recall} & \multicolumn{1}{l}{F1 score} \\
\midrule
Devign & 92.64& 27.68 & 17.1  & 21.14   & 94.36   & 24.46& 17.04 & 20.09   \\
Reveal & 76.85& 10.60  & \cellcolor{gray!25}40.03 & 16.68   & 76.16   & 5.53 & \cellcolor{gray!25}29.15 & 9.19\\
DeepWuKong & 86.36& 40.93 & 24.58 & 30.72   & 94.17   & 38.26& 10.27 & 16.19   \\
CodeBERT   & 95.75& 50.43 & 24.89 & 33.33   & \cellcolor{gray!25}96.61   & \cellcolor{gray!25}45.85& 13.04 & 20.30\\
CodeT5& \cellcolor{gray!25}97.21& \cellcolor{gray!25}55.67 & 30.42 & \cellcolor{gray!25}39.34   & 96.19   & 43.51& 24.55 & \cellcolor{gray!25}31.39   \\
EPVD   & \cellcolor{gray!45}98.10 & \cellcolor{gray!45}85.85 & \cellcolor{gray!45}78.77 & \cellcolor{gray!45}82.16   & \cellcolor{gray!45}98.10& \cellcolor{gray!45}88.17& \cellcolor{gray!45}76.30  & \cellcolor{gray!45}81.81  \\
\midrule
\tool & \cellcolor{gray!70}\textbf{99.04} & \cellcolor{gray!70}\textbf{90.90} & \cellcolor{gray!70}\textbf{86.14} & \cellcolor{gray!70}\textbf{88.45} & \cellcolor{gray!70}\textbf{98.97} & \cellcolor{gray!70}\textbf{90.90} & \cellcolor{gray!70}\textbf{84.47} & \cellcolor{gray!70}\textbf{87.56}\\
\bottomrule
\end{tabular}}
\label{tab:rq1}
\end{table*}
\subsection{RQ1. {Effectiveness of \tool in Original Setting}}

To answer RQ1, we conduct a comprehensive comparative analysis against the six vulnerability detection baseline methods across four performance metrics in the original setting, with results shown in Table~\ref{tab:rq1}. We can achieve the following observations.

\textbf{The proposed \tool consistently exhibits superior performance
contrasted with the baseline methods.} As delineated in Table~\ref{tab:rq1}, it shows that \tool surpasses all the baseline methods across all metrics, achieving the highest performance in each of the four metrics.
Specifically, \tool demonstrates 
improvements over the best baseline EPVD at
0.94\%, 5.05\%, 7.37\%, and 6.29\% across the four metrics, respectively.
The limited improvement of the accuracy metric could
be attributed to the inherent class imbalance within the dataset.
Compared with the average of previous vulnerability detection methods, \tool achieves an average absolute improvement of 45.71\%, 50.18\%, and 51.22\% in precision, recall and F1 score metrics, respectively.

\textbf{Pre-trained models perform better than GNN-based models.} We also discover that pre-trained models, such as CodeBERT, CodeT5, EPVD, and \tool consistently outperform GNN-based models across various performance indicators.
Specifically, these pre-trained models obtain 11 out of 12 Top-3 results, compared to only one for the GNN-based model.
When considering the average results, the four pre-training-based methods exhibit an average improvement of 12.24\%, 44.31\%, 27.82\%, and 37.97\% across the four metrics when compared to the three GNN-based methods, respectively. 
This observation highlights the advantages of the existing pre-trained models, which enhance the ability for capturing vulnerability patterns.
Furthermore, the combination of structural information
and semantic information (harnessed from
pre-trained models), as exemplified by EPVD and \tool, yielded the highest performance among all the vulnerability detection methods.

\begin{tcolorbox}
 \textbf{Answer to RQ1:} \tool outperforms all the baseline methods in all metrics, achieving
 0.94\%, 5.05\%, 7.37\%, and 6.29\% improvements over EPVD in Fan et al. dataset, respectively. Besides, the pre-trained models perform better than GNN-based models.
 \end{tcolorbox}

\subsection{RQ2. {Effectiveness of \tool in the
Identifier-Substitution Setting}}
To answer RQ2, we also compare \tool with the six vulnerability detection baseline methods in the identifier-substitution setting, in which user-defined variables and functions are mapped
to symbolic names. The results are shown in Table~\ref{tab:rq1}. We achieve the following findings.

\textbf{The performance shows most serious decline
among vulnerability detectors reliant solely on source code as input.} Specifically, CodeBERT and CodeT5 exclusively employ source code as their model input, and as illustrated in Table~\ref{tab:rq1} on the right column, their performance demonstrates a decline across seven out of eight evaluation cases. the F1 scores of CodeBERT and CodeT5 experience substantial reductions of -13.03\% and -7.95\%, respectively. DeepWukong also exhibits a substantial decrease in its F1 score, albeit with a simultaneous 7.81\% increase in accuracy. This suggests that following the substitution of user-defined identifier names, DeepWukong tends to classify samples as non-vulnerable, which is not conducive to effective vulnerability pattern detection.

\textbf{The simultaneous use of both structural paths and source code as inputs effectively mitigates the impact of identifier variations.} Both \tool and EPVD utilize a combination of structural paths and source code as inputs. Following user-defined identifier changes, these approaches exhibit minimal performance variations, with only marginal decreases of 0.89\% and 0.35\% in F1 scores, respectively.

Overall, it is evident that \tool emerges as the best-performing method, showcasing enhancements of 0.87\%, 2.73\%, 8.17\%, and 5.75\% in the four metrics respectively when compared to the best baseline methods. It emphasizes \tool's proficiency in implicitly capturing vulnerability-relevant patterns, even within an
identifier-substitution setting. Furthermore, it demonstrates the potential utility of incorporating game theory 
for software vulnerability detection, even in cases where the code substitutes use-defined identifiers.

\begin{tcolorbox}
 \textbf{Answer to RQ2:} The simultaneous use of both structural paths and source code as inputs effectively mitigates the impact of identifier variations. \tool consistently outperforms all baseline methods across all metrics, demonstrating enhancements over EPVD of 0.87\%, 2.73\%, 8.17\%, and 5.75\% in the identifier-substitution setting, respectively.
 \end{tcolorbox}

\subsection{RQ3. {Effectiveness of \tool in Vulnerability-Fix Pair Setting}}
\begin{table*}[htbp]
\centering

\setlength{\tabcolsep}{1.5mm}
\renewcommand{\arraystretch}{1.2}
\caption{
Comparison results between \tool and existing vulnerability detection methods in the vulnerability-fix pair setting (RQ3). The pair set concentrates on the vulnerable code and the corresponding fixed code, and the combine set encompasses all source code snippets.}
\resizebox{.97\textwidth}{!}{
\begin{tabular}{l|cccc@{\hspace{5mm}}|@{\hspace{5mm}}cccc}
\toprule
\diagbox{Metrics(\%) }{Setting}   & \multicolumn{4}{c|@{\hspace{5mm}}}{RQ3 (Pair)}  & \multicolumn{4}{c}{RQ3 (Combine)}  \\
\midrule
Method & Accuracy & \multicolumn{1}{l}{Precision} & \multicolumn{1}{l}{Recall} & \multicolumn{1}{l|@{\hspace{5mm}}}{F1 score} & \multicolumn{1}{l}{Accuracy} & \multicolumn{1}{l}{Precision} & \multicolumn{1}{l}{Recall} & \multicolumn{1}{l}{F1 score} \\
\midrule
Devign & 48.86& \cellcolor{gray!70}\textbf{57.76}  & 17.73  & 27.13& 93.59& 19.11 & 18.62  & 18.86\\
Reveal & 47.23& \cellcolor{gray!25}54.85  & 12.36  & 19.91& 84.45& 5.61  & 18.42  & 8.55 \\
DeepWuKong & 48.86& \cellcolor{gray!45}56.32  & 8.11   & 14.18& \cellcolor{gray!45}95.17& 26.45 & 9.34   & 13.80\\
CodeBERT   & \cellcolor{gray!25}51.37& 52.51  & 28.72  & 37.13& 93.55& \cellcolor{gray!25}36.24 & 28.72  & 32.05\\
CodeT5& \cellcolor{gray!45}52.27& 53.48  & \cellcolor{gray!25}34.98  & \cellcolor{gray!25}42.29& 92.86& 33.36 & \cellcolor{gray!25}34.98  & \cellcolor{gray!25}34.15\\
EPVD   & 50.81& 50.52  & \cellcolor{gray!45}78.77  & \cellcolor{gray!45}61.56& \cellcolor{gray!25}94.10& \cellcolor{gray!45}46.63 & \cellcolor{gray!45}78.77  & \cellcolor{gray!45}58.58\\
\midrule

\tool & \cellcolor{gray!70}\textbf{55.90} & 53.67 & \cellcolor{gray!70}\textbf{86.17} & \cellcolor{gray!70}\textbf{66.15} & \cellcolor{gray!70}\textbf{96.14} & \cellcolor{gray!70}\textbf{51.74} & \cellcolor{gray!70}\textbf{86.42} & \cellcolor{gray!70}\textbf{64.72}\\
\bottomrule
\end{tabular}}
\label{tab:RQ3}
\end{table*}
To answer RQ3, we compare \tool with the previous methods under vulnerability-fix pair setting. This setting is divided into two main sub-settings:
one that concentrates on the vulnerable code and corresponding fixed code (\ie Pair set), and another that encompasses all source code snippets (\ie Combine set). It presents a substantial challenge to distinguish whether a vulnerability has been effectively fixed or not in the setting. Typically, only a minor part of the code undergoes fixing, and the variable names within the code remain nearly identical between vulnerable code and corresponding fixed code.

As depicted in Table~\ref{tab:RQ3}, our evaluation compares the performance of \tool{} with six baseline vulnerability detection methods within the vulnerability-fix pair setting. The results demonstrate \tool{}'s best performance in 7 out of 8 cases.
For example, in the combined test set, \tool{} exhibits superiority by achieving improvements of 0.97\% in accuracy, 5.11\% in precision, 7.65\% in recall, and 6.14\% in F1 score compared to the best baseline methods. Notably, in the pair test set,  GNN-based methods excel in precision, although this advantage is coupled with sacrifices in accuracy, recall, and F1 score metrics.

Our experiment further underscores the potential of existing vulnerability detection approaches in distinguishing vulnerabilities and the corresponding fixed code.
This implies that current approaches frequently classify code samples before and after fixes as
the same label, potentially leading to
detrimental consequences in real-world scenarios. Once a developer commits the code snippet to rectify the identified vulnerability, the vulnerability detector is expected to accurately identify
the patched code segment. In fact, although \tool{} detects more than 52 pairs of the vulnerable code and corresponding fixed code when compared with best baseline methods, there remains substantial potential for enhancement.

\begin{tcolorbox}
 \textbf{Answer to RQ3:} In the vulnerability-fix pair setting, \tool performs better than the best baseline in most cases. \tool outperforms  0.97\%, 5.11\%, 7.65\%, and 6.14\% than EPVD in the combine set, respectively. 
 \end{tcolorbox}

\subsection{RQ4. {Effectiveness of \tool in Time-Split Setting}}
\begin{table}[htbp]
\centering
\begin{subtable}[h]{0.48\textwidth}
\setlength{\tabcolsep}{1.2mm}
\renewcommand{\arraystretch}{1.2}
\caption{Comparison results in the time-split setting.}
\resizebox{.99\textwidth}{!}{
\begin{tabular}{l|cccc}
\toprule
\diagbox{Metrics(\%) }{Setting}   & \multicolumn{4}{c|}{RQ4 (Time Split)}    \\
\midrule
Method & Accuracy & \multicolumn{1}{l}{Precision} & \multicolumn{1}{l}{Recall} & \multicolumn{1}{l|}{F1 score} \\
\midrule
Devign     & 92.44 & 26.54 & 18.53 & 21.82 \\
Reveal     & 79.16 & 10.69 & \cellcolor{gray!25}35.71 & 16.35 \\
DeepWuKong & 84.69 & \cellcolor{gray!25}36.05 & 26.03 & \cellcolor{gray!25}30.23 \\
CodeBERT   & 93.18 & 7.35  & 2.55  & 3.79  \\
CodeT5     & \cellcolor{gray!25}94.45 & 4.80  & 2.14  & 2.96  \\
EPVD       & \cellcolor{gray!45}96.63 & \cellcolor{gray!45}74.42 & \cellcolor{gray!45}68.83 & \cellcolor{gray!42}71.51 \\
\midrule
\tool & \cellcolor{gray!70}\textbf{97.48} & \cellcolor{gray!70}\textbf{76.83} & \cellcolor{gray!70}\textbf{74.69} & \cellcolor{gray!70}\textbf{75.75} \\
\bottomrule
\end{tabular}}
\end{subtable}
\hfill
\begin{subtable}[h]{0.48\textwidth}
\setlength{\tabcolsep}{1.2mm}
\renewcommand{\arraystretch}{1.2}
\caption{
The ablation study of \tool. }
\resizebox{.99\textwidth}{!}{
\begin{tabular}{c|l|cccc}
\toprule
Settings   & Module  & \multicolumn{1}{c}{Accuracy} & \multicolumn{1}{c}{Precision} & \multicolumn{1}{c}{Recall} & \multicolumn{1}{c}{F1 score} \\
\hline{}
\multirow{5}{*}{RQ1}    &\multirow{2}{*}{w/o Game}& 98.41   & 86.49    & 74.26 & 79.91   \\
& & ($\downarrow$0.63\%)	&($\downarrow$4.41\%)&	($\downarrow$11.88\%)&	($\downarrow$8.54\%)
\\ \cline{2-6}
   & \multirow{2}{*}{w/o Prototype}& 98.82&89.11&84.22&86.60    \\
& & ($\downarrow$0.22\%)	&($\downarrow$1.79\%)&	($\downarrow$1.92\%)	&($\downarrow$1.85\%)
    \\ \cline{2-6}
   & \tool & 99.04   & 90.90    & 86.14 & 88.45   \\
\midrule
\multirow{5}{*}{RQ2} &\multirow{2}{*}{w/o Game} &97.96   & 74.55    & 79.61 & 77.00   \\
& & ($\downarrow$1.01\%)	&($\downarrow$16.35\%)&	($\downarrow$4.86\%)&	($\downarrow$10.56\%)    \\ \cline{2-6}
   & \multirow{2}{*}{w/o Prototype}& 98.73&85.99&84.69&85.34   \\
   & & ($\downarrow$0.24\%)&($\downarrow$4.91\%) & ($\uparrow$-0.22\%)&($\downarrow$2.22\%)    \\ \cline{2-6}
   & \tool & 98.97   & 90.90    & 84.47 & 87.56  \\
\bottomrule
\end{tabular}}
\end{subtable}
\caption{
(a) Comparison results between \tool and vulnerability detection methods in the time-split setting (RQ4). 
The best result for each metric is highlighted in bold. 
(b) The impact of the zero-sum game construction (\ie Game) module and the class-level prototype learning (\ie Prototype) module on the performance of \tool (RQ5).}
\label{tab:rq4}
\end{table}

In RQ4, we explore a partitioning setting for the Fan et al. dataset based on time, which is motivated by Jimenez et al.~\cite{DBLP:conf/sigsoft/JimenezRPSTH19}. It dictates that at a specific point in time, denoted as $t_1$, only vulnerabilities known up to and including time $t_1$ should be available for training. Consequently, vulnerabilities discovered or made public after time $t_1$ should not be accessible for training prior to their appearance. Furthermore, we introduced a distinct time point $t_2$ to signify that data between $t_1$  and $t_2$ should be treated as the validation set, while data originating after $t_2$ is designated as the test set. This rigorous approach effectively mitigates the risk of data leakage. While the time-split setting underscores the imperative for robust vulnerability detection methods~\cite{DBLP:journals/ese/GargDJCPT22}, it necessitates that the vulnerability detection system possesses the capability to discern patterns associated with genuine vulnerabilities, without becoming overly fixated on semantics or irrelevant information to the vulnerabilities.

As delineated in Table~\ref{tab:rq4} (a), \tool continues to outperform other baseline methods. Specifically, under the time-split setting, \tool exhibits performance enhancements of 0.85\%, 2.41\%, 5.86\%, and 4.24\% across the four metrics, respectively. This underscores \tool's superior ability to recognize vulnerability-related information in real-world scenarios, surpassing its counterparts.

Table~\ref{tab:rq4} (a) also shows that baseline methods incorporating graph or path structures (\ie Devign, Reveal, DeepWuKong, EPVD, and \tool) consistently outperform those relying solely on sequences (\ie CodeBERT and CodeT5). It is noteworthy that methods utilizing a GNN-based model do not exhibit substantial performance degradation under the time-split setting, with performance variations of 0.68\%, -0.33\%, and -0.49\%, respectively. In contrast, while EPVD and \tool perform exceptionally well, there is still some performance degradation.

\begin{tcolorbox}
 \textbf{Answer to RQ4:} In the time-split setting, \tool outperforms 0.85\%, 2.41\%, 5.86\%, and 4.24\% than best baseline in the four metrics, respectively. The methods incorporating graph or path structures consistently outperform those relying solely on sequence methods in this real-world
 setting.
 \end{tcolorbox}

\subsection{RQ5. {Effectiveness of Different Modules in \tool}}

To answer RQ5, we explore the effectiveness of different modules on the performance of \tool. Specifically, we study the two involved modules including the Zero-sum Game Construction (\ie Game) module and Class-level Prototype Learning (\ie Prototype) module.

\subsubsection{Zero-sum Game Construction Module}

To understand the impact of this module, we deploy a variant of \tool without the zero-sum game construction module (\ie Game). It only directly uses the source code as a training process without constructing the Detector and Calibrator. 

Table~\ref{tab:rq4} (b) shows the performance of the variant on the two settings in the Fan et al. dataset. The addition of the Game module yields enhancements across all four metrics in the original and identifier-substitution settings, with an average improvement by 0.82\%, 10.38\%, 8.37\%, and 9.55\%, respectively. Specifically, the Game module improves the F1 score by 8.54\% and 10.56\%, respectively. It is worth mentioning that the Game module achieves a 16.35\% improvement in precision, 
which indicates that it can better recognize similar codes in the identifier-substitution setting.
Overall, the results show that the Game module can capture the semantic-agnostic features to enhance the vulnerability detection performance.
\subsubsection{Class-level Prototype Learning Module}
To explore the contribution of the class-level prototype learning module, we also construct a variant of \tool without the class-level prototype learning (\ie Prototype) module. The other setting of this variant is
consistent with \tool.

As shown in Table~\ref{tab:rq4} (b), this variant improves the \tool performance in 7 out of the 8 cases, which achieves average improvements of 0.23\%, 3.35\%, 0.85\% and 2.04\% across four metrics, respectively.  In the identifier-substitution setting, the class-level prototype learning module exhibits a lower performance improvement compared with the original setting. This can be attributed to that this difficult setting leads to a challenge in prototype learning, which makes it more difficult to accurately capture vulnerability patterns. 
\begin{tcolorbox}
 \textbf{Answer to RQ5:} 
The zero-sum game construction and class-level prototype learning module can improve the performance of \tool. The Game module has a greater effect compared with the Prototype module, which averagely improves the performance of 0.82\%, 10.38\%, 8.37\%, and 9.55\% across four metrics, respectively.
 \end{tcolorbox}

\subsection{RQ6. {Influence of Different Hyper-Parameters in \tool}}
To answer RQ6, we explore the impact of different hyper-parameters, including the training batch size and the weight of regularization loss $\mathcal{L}_{reg}$.
\begin{figure*}[t]
	\centering
	\includegraphics[width=1.0\textwidth]{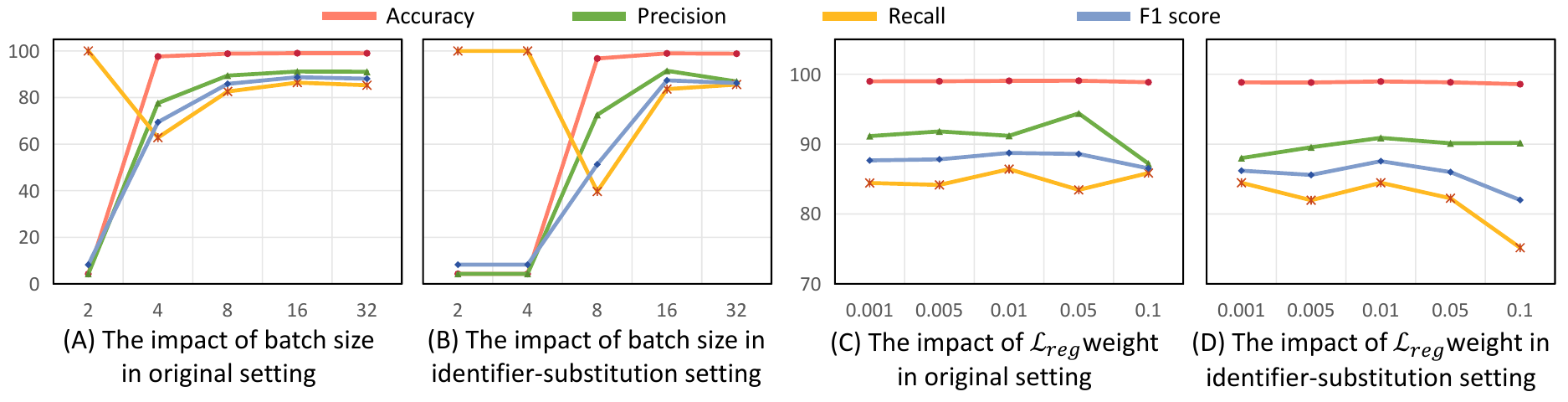}
    \caption{(A) and (B) respectively denote the performance impacts of varying batch sizes for \tool, within the original and identifier-substitution settings. (C) and (D) correspondingly denote the performance resulting from different weights of regularization loss in \tool. }
\label{fig:rq6}
\end{figure*}
\subsubsection{Batch Size}
In the zero-sum game construction module, we propose a calibrator $\mathcal{C}$ to distinguish the vulnerable code from the corresponding fixed code. We conduct the experiment on how the batch size (i.e., the pair of vulnerable code and the corresponding fixed codes) impacts the performance of \tool.

Fig.~\ref{fig:rq6} (A) and (B) show the performance of \tool across four metrics with different batch sizes in original and identifier-substitution settings, respectively. The larger batch sizes consistently yield improved performance for \tool. When the batch size is set to 16, we achieve performance of 99.06\%, 91.19\%, 86.42\%, and 88.74\% across the four metrics, respectively. However, for batch sizes exceeding 16, we observe that the \tool's performance exhibits relative stability.  It is difficult for the model to learn vulnerability-related information from the less varied utterances when the batch size is small. Furthermore, our analysis reveals that the identifier-substitution setting is more susceptible to batch size. Nevertheless, the trend of improved performance with larger batch sizes remains consistent across both original and identifier-substitution settings.

\subsubsection{Weight of Regularization Loss}
In the class-level prototype learning module, we propose a regularization loss $\mathcal{L}_{reg}$ to improve the generalization for vulnerability detection.
We conduct the experiment to explore the effect of regularization weight.

As shown in Fig.~\ref{fig:rq6} (C) and (D), the influence of regularization weight on \tool performance is relatively modest, exhibiting a discernible pattern of improvement followed by a decline. Notably, the optimal weight value, where performance is maximized, is found at 0.01. It is noteworthy that the regularization's impact on performance variance is more pronounced in a context where semantic relevance is absent. This observation underscores the efficacy of regularization weight in enhancing the ability to detect vulnerability patterns within semantically similar setting.

\begin{tcolorbox}
 \textbf{Answer to RQ6:} 
The performance of \tool is influenced by batch size and weight of regularization loss. Our default settings have undergone optimization to yield optimal results.
 \end{tcolorbox}

\section{Discussion}
\label{sec:discussion}

\subsection{Why does \tool Work Well?}
\begin{figure*}[t]
	\centering
	\includegraphics[width=0.90\textwidth]{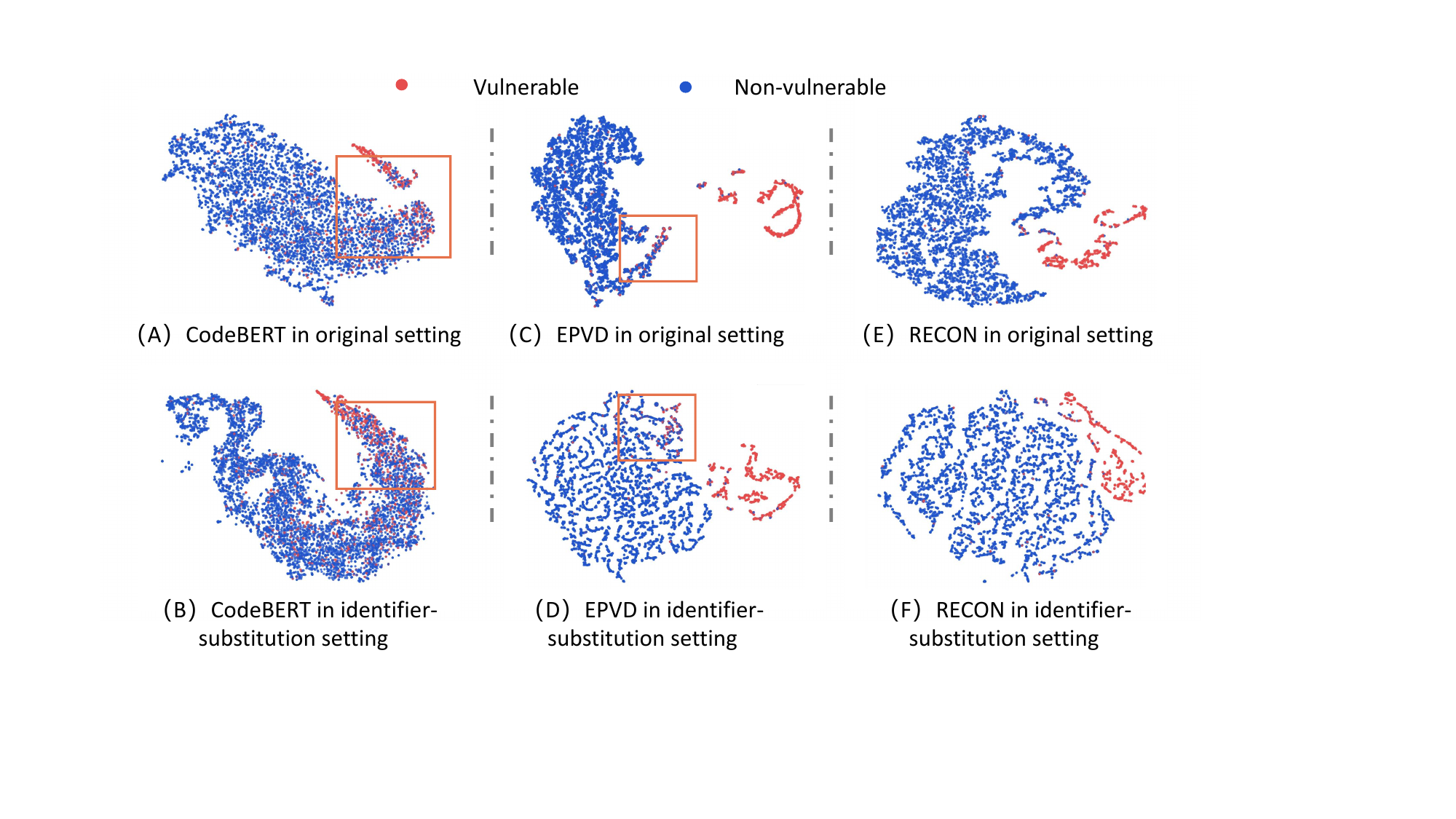}
    \caption{T-SNE visualization of CodeBERT in (A) and (B), T-SNE visualization of EPVD in (C) and (D), and T-SNE visualization of \tool in (E) and (F).
    The upper set of figures (i.e. (A), (C), and (E)) denote the original settings, while the lower set of figures (i.e. (B), (D), and (F)) illustrate identifier-substitution settings. The blue dots denote the vulnerable and the red dots represent the vulnerable code snippets.}
\label{fig:tsne}
\end{figure*}

\textbf{(1) The ability to learn the vulnerability representation in original and identifier-substitution settings.} The proposed zero-sum game construction module and class-level prototype learning module greatly
contribute to \tool by enabling it to learn vulnerability representations effectively. More specifically, we design a Calibrator to capture semantic-agnostic features, thereby improving the performance of the Detector. We employ the T-SNE~\cite{van2008visualizing/TSNE} technique for visualizing the vulnerability representations within CodeBERT (\ie Fig.~\ref{fig:tsne} (A) and (B)), EPVD (\ie Fig.~\ref{fig:tsne} (C) and (D)), and \tool (\ie Fig.~\ref{fig:tsne} (E) and (F)), 
in both the original and identifier-substitution settings.

As illustrated in Fig.~\ref{fig:tsne}, it becomes apparent that CodeBERT does not provide a discriminative vulnerability representation. Both EPVD and \tool exhibit an ability to discern vulnerability characteristics. However, it is noteworthy that EPVD's decision-making tends to be overconfidence. This is evidenced by the presence of the orange boxes in Fig.~\ref{fig:tsne} (C) and (D), indicating that EPVD inadvertently obfuscates certain semantically similar vulnerability samples.
The challenge posed by the identifier-substitution setting exacerbates this issue. Conversely, \tool demonstrates a discriminative ability to effectively distinguish between the representations.

\begin{figure*}[t]
	\centering
	\includegraphics[width=0.85\textwidth]{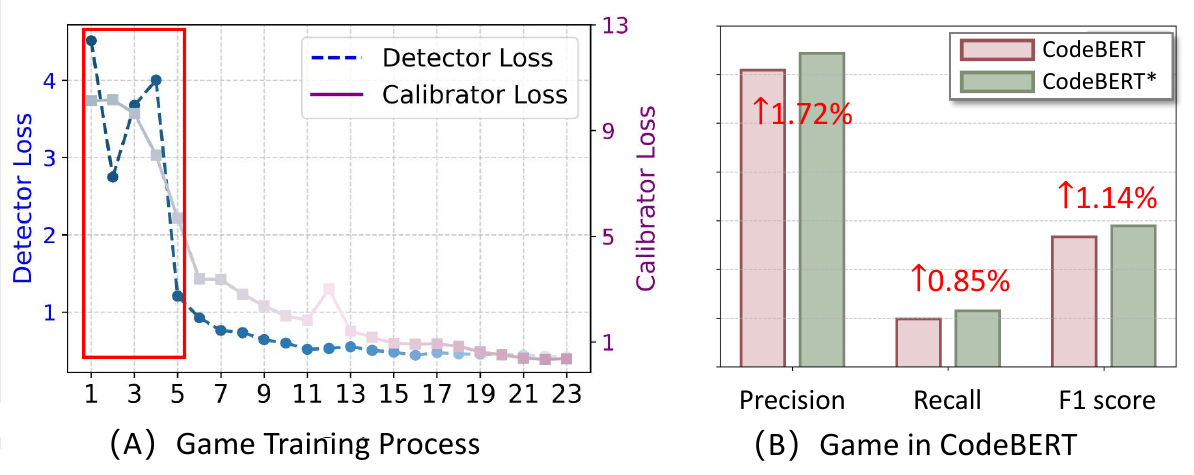}
    \caption{(A) The loss functions of both Detector and Calibrator throughout the game training process. (B) The performance between the CodeBERT and CodeBERT*.  The CodeBERT* is the CodeBERT improved by the zero-sum game. }
\label{fig:discussion}
\end{figure*}

\textbf{(2) The gaming process of \tool leads to an increase in the vulnerability detection performance.} The proposed zero-sum game construction module is developed based on a zero-sum game framework involving two key players: Detector and Calibrator, which enhances the overall performance of vulnerability detection.
Fig.~\ref{fig:discussion} (A)
visualizes the loss functions of both Detector and Calibrator throughout the training process.

As can be seen, during the training phase, the Detector and Calibrator engage in a strategic interaction. Initially, they compete with each other, akin to a zero-sum game in the early stage of the training process (highlighted in red box). During this phase, as one player's loss increases, the other player's loss also experiences a corresponding rise. This dynamic reflects the competitive nature of their interaction. During the latter training process, a noteworthy shift occurs, both Detector and Calibrator transition from competition to cooperation, resulting in a mutually beneficial outcome. In this process, the losses for both players simultaneously decrease. However, it is important to note that the rate of reduction differs between them. It shows that both players are working towards a shared objective (i.e. software vulnerability detection), which captures the semantic-agnostic features of the Calibrator for enhancing the Detector performance for vulnerability detection. 

\textbf{(3) The zero-sum game can be constructed in different methods.}
In this paper, we have identified the advantages of constructing the zero-sum game and prototype learning
for
software vulnerability detection. Another advantage of \tool is its flexibility and extensibility. \tool can be applied to the existing vulnerability detection methods.

For example, as shown in Fig.~\ref{fig:discussion} (B), we enhance the CodeBERT method
by incorporating the zero-sum game and prototype learning (\ie, CodeBERT*). The improved CodeBERT* performs better than the original CodeBERT by 1.72\%, 0.85\% and 1.14\% across the precision, recall, and F1 score metrics, respectively. This empirical observation further underscores the effectiveness of the proposed framework in vulnerability detection.


\subsection{Threats and Limitations}

\textbf{Dataset Validity Concerns}:
One potential threat to the validity of our study arises from the dataset we have constructed. In this paper, owing to constraints in dataset availability, we have exclusively relied upon Fan et al. dataset, albeit employing diverse segmentation techniques to mitigate potential data leakage. In our future work, we are committed to constructing a more representative benchmark dataset for more comprehensive evaluation.

\textbf{Programming Language Scope}:
Another threat pertains to the choice of programming languages considered. All instances of vulnerabilities in our paper have been developed using the C/C++ programming languages. As a result, we have not incorporated other widely used languages, such as Java and Python. However, \tool can be applied to the other programming languages. In future research, we intend to extend our investigations to these alternative programming languages.

\textbf{Baseline Implementation}:
The third threat is the implementation of baseline models. Throughout each experimental configuration, we try our best to utilize the original source code of the baseline models obtained from the GitHub repositories maintained by the respective authors. Furthermore, we have maintained consistency by employing the identical hyperparameters in the original papers authored.

\section{Related Work}
\label{sec:related}
Software vulnerability detection is a crucial component of software security, which involves the identification and mitigation of potential security risks. The traditional vulnerability detection methods are Program Analysis (PA)-based methods~\cite{redebug, DBLP:conf/IEEEares/LiKBL13/symbolic2, fuzz3,argos}, which necessitate expert knowledge and extract manual features tailored to specific vulnerability types, such as BufferOverflow~\cite{DBLP:conf/uss/LarochelleE01, DBLP:journals/spe/LheeC03}. 
In contrast, recent research has increasingly focused on learning-based software vulnerability detection, which offers the capacity to identify more vulnerability types and a greater quantity of vulnerabilities~\cite{vuldeepecker}. Moreover, these approaches have the ability to learn implicitly vulnerability patterns from historical data pertaining to known vulnerabilities~\cite{IVDETECT}.
Learning-based vulnerability detection techniques can be broadly classified into two categories based on the input representations and training model utilized: sequence-based methods and graph-based methods.

Sequence-based methods~\cite{8322752, DBLP:journals/corr/abs-1708-02368, app10051692, DBLP:conf/codaspy/GriecoGURFM16} convert code into token sequences, which first use the Recurrent Neural Networks (RNNs) and LSTM to learn the features. For example, VulDeepecker~\cite{vuldeepecker} generates the code gadgets from the source code as the granularity to train the Bidirectional (Bi)-LSTM network for vulnerability detection. SySeVR~\cite{sysevr} also extracts code gadgets by traversing AST and then leverages a Bi-LSTM network.
In recent years, with the increase in the number and type of software vulnerabilities, the adoption of pre-trained models with subsequent fine-tuning has emerged as an approach for vulnerability detection~\cite{DBLP:conf/sigsoft/WangYGP0L22/nomorefinetune}.
For example, CodeBERT~\cite{DBLP:conf/emnlp/FengGTDFGS0LJZ20/codebert} is an encoder-only pre-trained model rooted in the Roberta architecture, which is applied to the down-steam task of vulnerability detection following a fine-tuning process.
CodeT5~\cite{DBLP:conf/emnlp/0034WJH21/CodeT5} is an innovative Transformer-based model that conceptualizes the vulnerability detection task within the sequence-to-sequence framework.

Compared with the sequence-based methods, the graph-based methods provide more structural information and specialize in more fine-grained levels~\cite{DBLP:conf/dsaa/NguyenNXAKDJ22}. Specifically, Graph-based methods~\cite{DBLP:journals/infsof/CaoSBWL21, DBLP:conf/ictai/PhanNB17,DBLP:journals/tifs/WangYTTHFFBW21,DBLP:conf/ijcai/DuanWJRLYW19} generate the structural graph from source code and use Graph Neutral Networks (GNNs) for software vulnerability detection. 
For example, CPGVA~\cite{DBLP:conf/icait/WangZWXH18/CPGVA} combines the AST, CFG, and DFG and generates the code property graph (CPG) to vulnerability detection. 
IVDetect~\cite{IVDETECT} utilizes a Program Dependency Graph (PDG) and combines different information into the vulnerability representation. 
Cao et al.~\cite{cao2022mvd} utilize the PDG and propose MVD to detect fine-grained memory-related vulnerability.
LineVD~\cite{linevd} uses the Graph Attention Transformer (GAT), which characterizes the data and controls dependency information in statement-level vulnerability detection.

In this paper, we focus on the zero-sum game to present a novel framework, which involves a zero-sum construction module for capturing the semantic-agnostic features and a class-level prototype learning module for capturing representative vulnerability patterns.

\section{Conclusion}
\label{sec:conclusion}
In this paper, we propose a software vulnerability detection framework with a zero-sum game and prototype learning, named \tool.
It mainly consists of a zero-sum game construction module for capturing the semantic-agnostic features and a class-level prototype learning module for capturing representative vulnerability patterns. In addition, we also design a balance gap-based training strategy to ensure the stability of training process.
Extensive experiments in software vulnerability detection are conducted to evaluate
the effectiveness of \tool in the popular benchmark dataset. In the future, we intend to further evaluate \tool on a broader range of datasets and programming languages for software vulnerability detection.

\section*{Data availability}
Our source code and experimental data are available at: \textit{{\http}}

\bibliographystyle{IEEEtran}
\bibliography{IEEEabrv, Citation}
\end{document}